
%
%
%

\catcode`\@=11 


\newwrite\FIGUREwrite
\def\PSFIGURE#1{\immediate\openout\FIGUREwrite=#1 \PSFG@begin\PSFG@end}
{\catcode`\^^M=\active %
 \gdef\PSFG@begin{
\catcode`\%=11 
   \begingroup \catcode`\^^M=\active \let^^M=\relax}%
 \gdef\PSFG@end#1{\PSFG@@end #1^^M\PSFG@terminate \endgroup
    \catcode`\%=14 }
 \gdef\PSFG@@end#1^^M{\toks0={#1}\immediate\write\FIGUREwrite{\the\toks0}%
   \futurelet\N@XT\PSFG@test}%
 \gdef\PSFG@test{\ifx\N@XT\PSFG@terminate \let\N@XT=\relax%
       \else \let\N@XT=\PSFG@@end \fi \N@XT}%
}
\let\PSFG@terminate=\relax


\font\twelvebf=cmbx12   
\font\ninerm=cmr9
\font\ninesl=cmsl9

%
\def\nolabels{\def\wrlabeL##1{}\def\eqlabeL##1{}\def\reflabeL##1{}}
\def\writelabels{\def\wrlabeL##1{\leavevmode\vadjust{\rlap{\smash%
{\line{{\escapechar=` \hfill\rlap{\sevenrm\hskip.03in\string##1}}}}}}}%
\def\eqlabeL##1{{\escapechar-1\rlap{\sevenrm\hskip.05in\string##1}}}%
\def\reflabeL##1{\noexpand\llap{\noexpand\sevenrm\string\string\string##1}}}
\nolabels
%
\global\newcount\secno \global\secno=0
\global\newcount\meqno \global\meqno=1
\def\newsec#1{\global\advance\secno by1
\global\subsecno=0\eqnres@t\medskip\noindent{\it\the\secno. #1 ---}
\writetoca{{\secsym} {#1}}}
\def\eqnres@t{\xdef\secsym{\the\secno.}\global\meqno=1}
\def\sequentialequations{\def\eqnres@t{\bigbreak}}\xdef\secsym{}
\global\newcount\subsecno \global\subsecno=0
\def\subsec#1{\global\advance\subsecno by1\message{(\secsym\the\subsecno. #1)}
\ifnum\lastpenalty>9000\else\bigbreak\fi
\noindent{\it\secsym\the\subsecno. #1}\writetoca{\string\quad
{\secsym\the\subsecno.} {#1}}\par\nobreak\medskip\nobreak}
\def\appendix#1#2{\global\meqno=1\global\subsecno=0\xdef\secsym{\hbox{#1.}}
\bigbreak\bigskip\noindent{\bf Appendix #1. #2}\message{(#1. #2)}
\writetoca{Appendix {#1.} {#2}}\par\nobreak\medskip\nobreak}
%
%
\def\eqnn#1{\xdef #1{(\secsym\the\meqno)}\writedef{#1\leftbracket#1}%
\global\advance\meqno by1\wrlabeL#1}
\def\eqna#1{\xdef #1##1{\hbox{$(\secsym\the\meqno##1)$}}
\writedef{#1\numbersign1\leftbracket#1{\numbersign1}}%
\global\advance\meqno by1\wrlabeL{#1$\{\}$}}
\def\eqn#1#2{\xdef #1{(\secsym\the\meqno)}\writedef{#1\leftbracket#1}%
\global\advance\meqno by1$$#2\eqno#1\eqlabeL#1$$}
%
%
%
\global\newcount\refno \global\refno=1
\newwrite\rfile
\def\ref{[\the\refno]\nref}
\def\nref#1{\xdef#1{[\the\refno]}\writedef{#1\leftbracket#1}%
\ifnum\refno=1\immediate\openout\rfile=refs.tmp\fi
\global\advance\refno by1\chardef\wfile=\rfile\immediate
\write\rfile{\noexpand\item{$^{#1}$}\reflabeL{#1\hskip.31in}\pctsign}\findarg}
\def\findarg#1#{\begingroup\obeylines\newlinechar=`\^^M\pass@rg}
{\obeylines\gdef\pass@rg#1{\writ@line\relax #1^^M\hbox{}^^M}%
\gdef\writ@line#1^^M{\expandafter\toks0\expandafter{\striprel@x #1}%
\edef\next{\the\toks0}\ifx\next\em@rk\let\next=\endgroup\else\ifx\next\empty%
\else\immediate\write\wfile{\the\toks0}\fi\let\next=\writ@line\fi\next\relax}}
\def\striprel@x#1{} \def\em@rk{\hbox{}}
\def\lref{\begingroup\obeylines\lr@f}
\def\lr@f#1#2{\gdef#1{\ref#1{#2}}\endgroup\unskip}
\def\semi{;\hfil\break}
\def\addref#1{\immediate\write\rfile{\noexpand\item{}#1}} 
\def\startrefs#1{\immediate\openout\rfile=refs.tmp\refno=#1}
\def\xref{\expandafter\xr@f}\def\xr@f[#1]{#1}
\def\refs#1{\count255=1[\r@fs #1{\hbox{}}]}
\def\r@fs#1{\ifx\und@fined#1\message{reflabel \string#1 is undefined.}%
\nref#1{need to supply reference \string#1.}\fi%
\vphantom{\hphantom{#1}}\edef\next{#1}\ifx\next\em@rk\def\next{}%
\else\ifx\next#1\ifodd\count255\relax\xref#1\count255=0\fi%
\else#1\count255=1\fi\let\next=\r@fs\fi\next}
%

%
\newwrite\ffile\global\newcount\figno \global\figno=1
\def\fig{fig.~\the\figno\nfig}
\def\nfig#1{\xdef#1{fig.~\the\figno}%
\writedef{#1\leftbracket fig.\noexpand~\the\figno}%
\ifnum\figno=1\immediate\openout\ffile=figs.tmp\fi\chardef\wfile=\ffile%
\immediate\write\ffile{\noexpand\medskip\noexpand\item{Fig.\ \the\figno. }
\reflabeL{#1\hskip.55in}\pctsign}\global\advance\figno by1\findarg}
\def\xfig{\expandafter\xf@g}\def\xf@g fig.\penalty\@M\ {}
\def\figs#1{figs.~\f@gs #1{\hbox{}}}
\def\f@gs#1{\edef\next{#1}\ifx\next\em@rk\def\next{}\else
\ifx\next#1\xfig #1\else#1\fi\let\next=\f@gs\fi\next}
\newwrite\lfile
{\escapechar-1\xdef\pctsign{\string\%}\xdef\leftbracket{\string\{}
\xdef\rightbracket{\string\}}\xdef\numbersign{\string\#}}

\def\writestop{\def\writestoppt{\immediate\write\lfile{\string\pageno%
\the\pageno\string\startrefs\leftbracket\the\refno\rightbracket%
\string\def\string\secsym\leftbracket\secsym\rightbracket%
\string\secno\the\secno\string\meqno\the\meqno}\immediate\closeout\lfile}}
\def\writestoppt{}\def\writedef#1{}
\def\seclab#1{\xdef #1{\the\secno}\writedef{#1\leftbracket#1}\wrlabeL{#1=#1}}
\def\subseclab#1{\xdef #1{\secsym\the\subsecno}%
\writedef{#1\leftbracket#1}\wrlabeL{#1=#1}}
\newwrite\tfile \def\writetoca#1{}
\def\leaderfill{\leaders\hbox to 1em{\hss.\hss}\hfill}
\def\writetoc{\immediate\openout\tfile=toc.tmp
   \def\writetoca##1{{\edef\next{\write\tfile{\noindent ##1
   \string\leaderfill {\noexpand\number\pageno} \par}}\next}}}

%


\ \tolerance=10000
\parindent=10pt
\parskip=0pt
\def\today{\ifcase\month\or
   January\or February\or March\or April\or May\or June\or
   July\or August\or September\or October\or November\or December\fi
   \space\number\day, \number\year}
\newdimen\pagewidth \newdimen\pageheight \newdimen\textheight
\newdimen\headheight \newdimen\footheight
\hsize=7.0truein
\hoffset=-0.25truein
\vsize=10.2truein 
\voffset=-0.6truein
\baselineskip=12pt
\pagewidth=\hsize \pageheight=\vsize
\headheight=32pt
\textheight=672pt 
\footheight=20pt
\maxdepth=2pt

\font\largeheadfont=cmcsc10
\headline={\hbox to\pagewidth{%
{\largeheadfont Brown University}
\hfil{\largeheadfont HET-851}}}
\nopagenumbers 
\footline={\hbox to\pagewidth{\hskip125pt%
Submitted to Physical Review Letters%
\hfil\folio}}
\def\evenpage{\hbox to\pagewidth{\folio\hfil}}
\def\oddpage{\hbox to\pagewidth{\hfil\folio}}
\def\onepageout#1{\shipout\vbox{
      \offinterlineskip 
      \vbox to\headheight{\the\headline \vskip10pt \hrule height1pt \vfill}
      \vbox to\textheight{#1} 
      \vbox to\footheight{\vfil
              \ifnum\pageno=1\the\footline \else{\ifodd\pageno\oddpage
              \else\evenpage \fi}\fi}}
      \advancepageno}
\newbox\partialpage
\def\begindoublecolumns{\begingroup
       \output={\global\setbox\partialpage=\vbox{\unvbox255\bigskip}}\eject
       \output={\doublecolumnout} \hsize=3.375truein \vsize=1356pt}
\def\enddoublecolumns{\output={\balancecolumns}\eject
       \endgroup \pagegoal=\vsize}
\def\doublecolumnout{\splittopskip=\topskip \splitmaxdepth=\maxdepth
       \dimen@=672pt \advance\dimen@ by-\ht\partialpage
       \setbox0=\vsplit255 to\dimen@ \setbox2=\vsplit255 to\dimen@
       \onepageout\pagesofar \unvbox255 \penalty\outputpenalty}
\def\pagesofar{\unvbox\partialpage
       \wd0=\hsize \wd2=\hsize \hbox to\pagewidth{\box0\hfil\box2}}
\def\balancecolumns{\setbox0=\vbox{\unvbox255} \dimen@=\ht0
       \advance\dimen@ by\topskip \advance\dimen@ by-\baselineskip
       \divide\dimen@ by2 \splittopskip=\topskip
       {\vbadness=10000 \loop \global\setbox3=\copy0
           \global\setbox1=\vsplit3 to\dimen@
           \ifdim\ht3>\dimen@ \global\advance\dimen@ by1pt \repeat}
       \setbox0=\vbox to\dimen@{\unvbox1} \setbox2=\vbox to\dimen@{\unvbox3}
       \pagesofar}

\catcode`\@=12 


\PSFIGURE{HM1.ps}{
/mathworks 50 dict begin
/bdef {bind def} bind def
/xdef {exch def} bdef
/pgsv () def
/bpage {/pgsv save def} bdef
/epage {pgsv restore} bdef
/bplot {gsave} bdef
/eplot {grestore} bdef
/dx 0 def
/dy 0 def
/sides {/dx urx llx sub def /dy ury lly sub def} bdef
/llx 0 def
/lly 0 def
/urx 0 def
/ury 0 def
/bbox {/ury xdef /urx xdef /lly xdef /llx xdef sides} bdef
/por true def
/portrait {/por true def} bdef
/landscape {/por false def} bdef
/px 8.5 72 mul def
/py 11.0 72 mul def
/port {dx py div dy px div scale} bdef
/land {-90.0 rotate dy neg 0 translate dy py div dx px div scale} bdef
/csm {llx lly translate por {port} {land} ifelse} bdef
/SO { []        0 setdash } bdef
/DO { [0 4]     0 setdash } bdef
/DA { [4]       0 setdash } bdef
/DD { [0 4 3 4] 0 setdash } bdef
/M {moveto}  bdef
/L {{lineto} repeat stroke} bdef
/font_spec () def
/lfont currentfont def
/sfont currentfont def
/selfont {/font_spec xdef} bdef
/savefont {font_spec findfont exch scalefont def} bdef
/LF {lfont setfont} bdef
/SF {sfont setfont} bdef
/sh {show} bdef
/csh {dup stringwidth pop 2 div neg 0 rmoveto show} bdef
/rsh {dup stringwidth pop neg 0 rmoveto show} bdef
/r90sh {gsave currentpoint translate 90 rotate csh grestore} bdef
currentdict end def 
mathworks begin
/Times-Roman selfont
/lfont 20 savefont
/sfont 14 savefont
.5 setlinewidth 1 setlinecap 1 setlinejoin
end
mathworks begin
bpage
bplot
80 407 532 756 bbox portrait csm
SO
 78.09  77.33 757.00  77.33 757.00 570.67  78.09 570.67  78.09  77.33 M 4 L
LF
 73.09  71.33 M (0.6) rsh
 78.09 139.00  84.83 139.00 M 1 L
750.27 139.00 757.00 139.00 M 1 L
 73.09 133.00 M (0.8) rsh
 78.09 200.67  84.83 200.67 M 1 L
750.27 200.67 757.00 200.67 M 1 L
 73.09 194.67 M (1) rsh
 78.09 262.33  84.83 262.33 M 1 L
750.27 262.33 757.00 262.33 M 1 L
 73.09 256.33 M (1.2) rsh
 78.09 324.00  84.83 324.00 M 1 L
750.27 324.00 757.00 324.00 M 1 L
 73.09 318.00 M (1.4) rsh
 78.09 385.67  84.83 385.67 M 1 L
750.27 385.67 757.00 385.67 M 1 L
 73.09 379.67 M (1.6) rsh
 78.09 447.33  84.83 447.33 M 1 L
750.27 447.33 757.00 447.33 M 1 L
 73.09 441.33 M (1.8) rsh
 78.09 509.00  84.83 509.00 M 1 L
750.27 509.00 757.00 509.00 M 1 L
 73.09 503.00 M (2) rsh
 73.09 564.67 M (2.2) rsh
SF
129.19  77.33 129.19  82.53 M 1 L
129.19 565.47 129.19 570.67 M 1 L
159.07  77.33 159.07  82.53 M 1 L
159.07 565.47 159.07 570.67 M 1 L
180.28  77.33 180.28  82.53 M 1 L
180.28 565.47 180.28 570.67 M 1 L
196.73  77.33 196.73  82.53 M 1 L
196.73 565.47 196.73 570.67 M 1 L
210.17  77.33 210.17  82.53 M 1 L
210.17 565.47 210.17 570.67 M 1 L
221.53  77.33 221.53  82.53 M 1 L
221.53 565.47 221.53 570.67 M 1 L
231.37  77.33 231.37  82.53 M 1 L
231.37 565.47 231.37 570.67 M 1 L
240.05  77.33 240.05  82.53 M 1 L
240.05 565.47 240.05 570.67 M 1 L
LF
 78.09  55.33 M (10) rsh
SF
 79.09  60.13 M (-1) sh
247.82  77.33 247.82  82.53 M 1 L
247.82 565.47 247.82 570.67 M 1 L
298.91  77.33 298.91  82.53 M 1 L
298.91 565.47 298.91 570.67 M 1 L
328.80  77.33 328.80  82.53 M 1 L
328.80 565.47 328.80 570.67 M 1 L
350.01  77.33 350.01  82.53 M 1 L
350.01 565.47 350.01 570.67 M 1 L
366.45  77.33 366.45  82.53 M 1 L
366.45 565.47 366.45 570.67 M 1 L
379.89  77.33 379.89  82.53 M 1 L
379.89 565.47 379.89 570.67 M 1 L
391.26  77.33 391.26  82.53 M 1 L
391.26 565.47 391.26 570.67 M 1 L
401.10  77.33 401.10  82.53 M 1 L
401.10 565.47 401.10 570.67 M 1 L
409.78  77.33 409.78  82.53 M 1 L
409.78 565.47 409.78 570.67 M 1 L
LF
247.82  55.33 M (10) rsh
SF
248.82  60.13 M (0) sh
417.55  77.33 417.55  82.53 M 1 L
417.55 565.47 417.55 570.67 M 1 L
468.64  77.33 468.64  82.53 M 1 L
468.64 565.47 468.64 570.67 M 1 L
498.53  77.33 498.53  82.53 M 1 L
498.53 565.47 498.53 570.67 M 1 L
519.73  77.33 519.73  82.53 M 1 L
519.73 565.47 519.73 570.67 M 1 L
536.18  77.33 536.18  82.53 M 1 L
536.18 565.47 536.18 570.67 M 1 L
549.62  77.33 549.62  82.53 M 1 L
549.62 565.47 549.62 570.67 M 1 L
560.98  77.33 560.98  82.53 M 1 L
560.98 565.47 560.98 570.67 M 1 L
570.83  77.33 570.83  82.53 M 1 L
570.83 565.47 570.83 570.67 M 1 L
579.51  77.33 579.51  82.53 M 1 L
579.51 565.47 579.51 570.67 M 1 L
LF
417.55  55.33 M (10) rsh
SF
418.55  60.13 M (1) sh
587.27  77.33 587.27  82.53 M 1 L
587.27 565.47 587.27 570.67 M 1 L
638.37  77.33 638.37  82.53 M 1 L
638.37 565.47 638.37 570.67 M 1 L
668.25  77.33 668.25  82.53 M 1 L
668.25 565.47 668.25 570.67 M 1 L
689.46  77.33 689.46  82.53 M 1 L
689.46 565.47 689.46 570.67 M 1 L
705.91  77.33 705.91  82.53 M 1 L
705.91 565.47 705.91 570.67 M 1 L
719.35  77.33 719.35  82.53 M 1 L
719.35 565.47 719.35 570.67 M 1 L
730.71  77.33 730.71  82.53 M 1 L
730.71 565.47 730.71 570.67 M 1 L
740.55  77.33 740.55  82.53 M 1 L
740.55 565.47 740.55 570.67 M 1 L
749.23  77.33 749.23  82.53 M 1 L
749.23 565.47 749.23 570.67 M 1 L
LF
587.27  55.33 M (10) rsh
SF
588.27  60.13 M (2) sh
LF
757.00  55.33 M (10) rsh
SF
758.00  60.13 M (3) sh
203.45 114.91 288.31 259.12 345.08 356.60 366.45 387.55 443.09 463.43
527.96 500.89 612.82 514.16 697.68 518.19
M 7 L
200.15 111.51 M (o) sh  285.01 255.72 M (o) sh
341.78 353.20 M (o) sh  363.15 384.15 M (o) sh
439.79 460.03 M (o) sh  524.66 497.49 M (o) sh
609.52 510.76 M (o) sh  694.38 514.79 M (o) sh
LF
eplot
epage
end
showpage
}

\PSFIGURE{HM2.ps}{
/mathworks 50 dict begin
/bdef {bind def} bind def
/xdef {exch def} bdef
/pgsv () def
/bpage {/pgsv save def} bdef
/epage {pgsv restore} bdef
/bplot {gsave} bdef
/eplot {grestore} bdef
/dx 0 def
/dy 0 def
/sides {/dx urx llx sub def /dy ury lly sub def} bdef
/llx 0 def
/lly 0 def
/urx 0 def
/ury 0 def
/bbox {/ury xdef /urx xdef /lly xdef /llx xdef sides} bdef
/por true def
/portrait {/por true def} bdef
/landscape {/por false def} bdef
/px 8.5 72 mul def
/py 11.0 72 mul def
/port {dx py div dy px div scale} bdef
/land {-90.0 rotate dy neg 0 translate dy py div dx px div scale} bdef
/csm {llx lly translate por {port} {land} ifelse} bdef
/SO { []        0 setdash } bdef
/DO { [0 4]     0 setdash } bdef
/DA { [4]       0 setdash } bdef
/DD { [0 4 3 4] 0 setdash } bdef
/M {moveto}  bdef
/L {{lineto} repeat stroke} bdef
/font_spec () def
/lfont currentfont def
/sfont currentfont def
/selfont {/font_spec xdef} bdef
/savefont {font_spec findfont exch scalefont def} bdef
/LF {lfont setfont} bdef
/SF {sfont setfont} bdef
/sh {show} bdef
/csh {dup stringwidth pop 2 div neg 0 rmoveto show} bdef
/rsh {dup stringwidth pop neg 0 rmoveto show} bdef
/r90sh {gsave currentpoint translate 90 rotate csh grestore} bdef
currentdict end def 
mathworks begin
/Times-Roman selfont
/lfont 20 savefont
/sfont 14 savefont
.5 setlinewidth 1 setlinecap 1 setlinejoin
end
mathworks begin
bpage
bplot
80 407 532 756 bbox portrait csm
SO
 78.09  77.33 757.00  77.33 757.00 570.67  78.09 570.67  78.09  77.33 M 4 L
LF
 73.09  71.33 M (0.2) rsh
 78.09 139.00  84.83 139.00 M 1 L
750.27 139.00 757.00 139.00 M 1 L
 73.09 133.00 M (0.3) rsh
 78.09 200.67  84.83 200.67 M 1 L
750.27 200.67 757.00 200.67 M 1 L
 73.09 194.67 M (0.4) rsh
 78.09 262.33  84.83 262.33 M 1 L
750.27 262.33 757.00 262.33 M 1 L
 73.09 256.33 M (0.5) rsh
 78.09 324.00  84.83 324.00 M 1 L
750.27 324.00 757.00 324.00 M 1 L
 73.09 318.00 M (0.6) rsh
 78.09 385.67  84.83 385.67 M 1 L
750.27 385.67 757.00 385.67 M 1 L
 73.09 379.67 M (0.7) rsh
 78.09 447.33  84.83 447.33 M 1 L
750.27 447.33 757.00 447.33 M 1 L
 73.09 441.33 M (0.8) rsh
 78.09 509.00  84.83 509.00 M 1 L
750.27 509.00 757.00 509.00 M 1 L
 73.09 503.00 M (0.9) rsh
 73.09 564.67 M (1) rsh
 78.09  55.33 M (1) csh
213.87  77.33 213.87  82.53 M 1 L
213.87 565.47 213.87 570.67 M 1 L
213.87  55.33 M (1.5) csh
349.66  77.33 349.66  82.53 M 1 L
349.66 565.47 349.66 570.67 M 1 L
349.66  55.33 M (2) csh
485.44  77.33 485.44  82.53 M 1 L
485.44 565.47 485.44 570.67 M 1 L
485.44  55.33 M (2.5) csh
621.22  77.33 621.22  82.53 M 1 L
621.22 565.47 621.22 570.67 M 1 L
621.22  55.33 M (3) csh
757.00  55.33 M (3.5) csh
 78.09 570.67 145.98 463.79 213.87 384.94 281.77 324.78 349.66 277.63
417.55 239.86 485.44 209.04 553.33 183.48 621.22 162.01 689.11 143.75
757.00 128.07
M 10 L
DA
 78.09 570.67 145.98 459.81 213.87 379.39 281.77 318.69 349.66 271.53
417.55 234.00 485.44 203.54 553.33 178.39 621.22 157.34 689.11 139.50
757.00 124.21
M 10 L
DD
 78.09 570.67 145.98 448.25 213.87 362.95 281.77 300.51 349.66 253.08
417.55 215.98 485.44 186.26 553.33 162.00 621.22 141.86 689.11 124.91
757.00 110.47
M 10 L
eplot
epage
end
showpage
}

\PSFIGURE{HM3.ps}{
/mathworks 50 dict begin
/bdef {bind def} bind def
/xdef {exch def} bdef
/pgsv () def
/bpage {/pgsv save def} bdef
/epage {pgsv restore} bdef
/bplot {gsave} bdef
/eplot {grestore} bdef
/dx 0 def
/dy 0 def
/sides {/dx urx llx sub def /dy ury lly sub def} bdef
/llx 0 def
/lly 0 def
/urx 0 def
/ury 0 def
/bbox {/ury xdef /urx xdef /lly xdef /llx xdef sides} bdef
/por true def
/portrait {/por true def} bdef
/landscape {/por false def} bdef
/px 8.5 72 mul def
/py 11.0 72 mul def
/port {dx py div dy px div scale} bdef
/land {-90.0 rotate dy neg 0 translate dy py div dx px div scale} bdef
/csm {llx lly translate por {port} {land} ifelse} bdef
/SO { []        0 setdash } bdef
/DO { [0 4]     0 setdash } bdef
/DA { [4]       0 setdash } bdef
/DD { [0 4 3 4] 0 setdash } bdef
/M {moveto}  bdef
/L {{lineto} repeat stroke} bdef
/font_spec () def
/lfont currentfont def
/sfont currentfont def
/selfont {/font_spec xdef} bdef
/savefont {font_spec findfont exch scalefont def} bdef
/LF {lfont setfont} bdef
/SF {sfont setfont} bdef
/sh {show} bdef
/csh {dup stringwidth pop 2 div neg 0 rmoveto show} bdef
/rsh {dup stringwidth pop neg 0 rmoveto show} bdef
/r90sh {gsave currentpoint translate 90 rotate csh grestore} bdef
currentdict end def 
mathworks begin
/Times-Roman selfont
/lfont 20 savefont
/sfont 14 savefont
.5 setlinewidth 1 setlinecap 1 setlinejoin
end
mathworks begin
bpage
bplot
80 407 532 756 bbox portrait csm
SO
 78.09  77.33 757.00  77.33 757.00 570.67  78.09 570.67  78.09  77.33 M 4 L
LF
SF
 78.09 114.46  84.83 114.46 M 1 L
750.27 114.46 757.00 114.46 M 1 L
 78.09 136.18  84.83 136.18 M 1 L
750.27 136.18 757.00 136.18 M 1 L
 78.09 151.59  84.83 151.59 M 1 L
750.27 151.59 757.00 151.59 M 1 L
 78.09 163.54  84.83 163.54 M 1 L
750.27 163.54 757.00 163.54 M 1 L
 78.09 173.30  84.83 173.30 M 1 L
750.27 173.30 757.00 173.30 M 1 L
 78.09 181.56  84.83 181.56 M 1 L
750.27 181.56 757.00 181.56 M 1 L
 78.09 188.71  84.83 188.71 M 1 L
750.27 188.71 757.00 188.71 M 1 L
 78.09 195.02  84.83 195.02 M 1 L
750.27 195.02 757.00 195.02 M 1 L
LF
 59.09  71.33 M (10) rsh
SF
 60.09  76.13 M (-5) sh
 78.09 200.67  84.83 200.67 M 1 L
750.27 200.67 757.00 200.67 M 1 L
 78.09 237.79  84.83 237.79 M 1 L
750.27 237.79 757.00 237.79 M 1 L
 78.09 259.51  84.83 259.51 M 1 L
750.27 259.51 757.00 259.51 M 1 L
 78.09 274.92  84.83 274.92 M 1 L
750.27 274.92 757.00 274.92 M 1 L
 78.09 286.87  84.83 286.87 M 1 L
750.27 286.87 757.00 286.87 M 1 L
 78.09 296.64  84.83 296.64 M 1 L
750.27 296.64 757.00 296.64 M 1 L
 78.09 304.90  84.83 304.90 M 1 L
750.27 304.90 757.00 304.90 M 1 L
 78.09 312.05  84.83 312.05 M 1 L
750.27 312.05 757.00 312.05 M 1 L
 78.09 318.36  84.83 318.36 M 1 L
750.27 318.36 757.00 318.36 M 1 L
LF
 59.09 194.67 M (10) rsh
SF
 60.09 199.47 M (-4) sh
 78.09 324.00  84.83 324.00 M 1 L
750.27 324.00 757.00 324.00 M 1 L
 78.09 361.13  84.83 361.13 M 1 L
750.27 361.13 757.00 361.13 M 1 L
 78.09 382.85  84.83 382.85 M 1 L
750.27 382.85 757.00 382.85 M 1 L
 78.09 398.25  84.83 398.25 M 1 L
750.27 398.25 757.00 398.25 M 1 L
 78.09 410.21  84.83 410.21 M 1 L
750.27 410.21 757.00 410.21 M 1 L
 78.09 419.97  84.83 419.97 M 1 L
750.27 419.97 757.00 419.97 M 1 L
 78.09 428.23  84.83 428.23 M 1 L
750.27 428.23 757.00 428.23 M 1 L
 78.09 435.38  84.83 435.38 M 1 L
750.27 435.38 757.00 435.38 M 1 L
 78.09 441.69  84.83 441.69 M 1 L
750.27 441.69 757.00 441.69 M 1 L
LF
 59.09 318.00 M (10) rsh
SF
 60.09 322.80 M (-3) sh
 78.09 447.33  84.83 447.33 M 1 L
750.27 447.33 757.00 447.33 M 1 L
 78.09 484.46  84.83 484.46 M 1 L
750.27 484.46 757.00 484.46 M 1 L
 78.09 506.18  84.83 506.18 M 1 L
750.27 506.18 757.00 506.18 M 1 L
 78.09 521.59  84.83 521.59 M 1 L
750.27 521.59 757.00 521.59 M 1 L
 78.09 533.54  84.83 533.54 M 1 L
750.27 533.54 757.00 533.54 M 1 L
 78.09 543.31  84.83 543.31 M 1 L
750.27 543.31 757.00 543.31 M 1 L
 78.09 551.56  84.83 551.56 M 1 L
750.27 551.56 757.00 551.56 M 1 L
 78.09 558.72  84.83 558.72 M 1 L
750.27 558.72 757.00 558.72 M 1 L
 78.09 565.03  84.83 565.03 M 1 L
750.27 565.03 757.00 565.03 M 1 L
LF
 59.09 441.33 M (10) rsh
SF
 60.09 446.13 M (-2) sh
LF
 59.09 564.67 M (10) rsh
SF
 60.09 569.47 M (-1) sh
LF
SF
180.28  77.33 180.28  82.53 M 1 L
180.28 565.47 180.28 570.67 M 1 L
240.05  77.33 240.05  82.53 M 1 L
240.05 565.47 240.05 570.67 M 1 L
282.46  77.33 282.46  82.53 M 1 L
282.46 565.47 282.46 570.67 M 1 L
315.36  77.33 315.36  82.53 M 1 L
315.36 565.47 315.36 570.67 M 1 L
342.24  77.33 342.24  82.53 M 1 L
342.24 565.47 342.24 570.67 M 1 L
364.96  77.33 364.96  82.53 M 1 L
364.96 565.47 364.96 570.67 M 1 L
384.65  77.33 384.65  82.53 M 1 L
384.65 565.47 384.65 570.67 M 1 L
402.01  77.33 402.01  82.53 M 1 L
402.01 565.47 402.01 570.67 M 1 L
LF
 78.09  55.33 M (10) rsh
SF
 79.09  60.13 M (0) sh
417.55  77.33 417.55  82.53 M 1 L
417.55 565.47 417.55 570.67 M 1 L
519.73  77.33 519.73  82.53 M 1 L
519.73 565.47 519.73 570.67 M 1 L
579.51  77.33 579.51  82.53 M 1 L
579.51 565.47 579.51 570.67 M 1 L
621.92  77.33 621.92  82.53 M 1 L
621.92 565.47 621.92 570.67 M 1 L
654.81  77.33 654.81  82.53 M 1 L
654.81 565.47 654.81 570.67 M 1 L
681.69  77.33 681.69  82.53 M 1 L
681.69 565.47 681.69 570.67 M 1 L
704.42  77.33 704.42  82.53 M 1 L
704.42 565.47 704.42 570.67 M 1 L
724.10  77.33 724.10  82.53 M 1 L
724.10 565.47 724.10 570.67 M 1 L
741.47  77.33 741.47  82.53 M 1 L
741.47 565.47 741.47 570.67 M 1 L
LF
417.55  55.33 M (10) rsh
SF
418.55  60.13 M (1) sh
LF
757.00  55.33 M (10) rsh
SF
758.00  60.13 M (2) sh
638.37 178.85 468.64 309.31 315.36 420.62 272.62 450.79 159.07 522.66 M 4 L
635.07 175.45 M (o) sh  465.34 305.91 M (o) sh
312.06 417.22 M (o) sh  269.32 447.39 M (o) sh
155.77 519.26 M (o) sh
LF
eplot
epage
end
showpage
}


\def\Title#1#2{\centerline{\twelvebf #2}\bigskip}

\def\authornote#1#2{{#1}}
\def\footnote#1#2{\ref\NONE{#2}} 
\def\INSERTFIG#1{#1}

\def\ABSTRACT#1{
{\ninerm
\centerline{(\DATE)}
\smallskip \parindent=.75in
{\narrower \parindent=10pt
{#1}
\medskip}}
\parindent=10pt
\begindoublecolumns}



\def\OMIT#1{}
\def\semi{; }
\def\MU{\mu}
\def\vev#1{\langle #1 \rangle}

\def\etal{{\it et al.\/}}
\def\ccdot{\hbox{\kern-.1em$\cdot$\kern-.1em}}
\def\vv{v \ccdot v'}

\def\dt{{\rm d}t}
\def\pvint{-\mskip-15mu\int}
\def\REF{\nref}
\def\nextline{}
\def\art#1{{\sl #1}}
\def\art#1{}
\def\NP{{\it Nucl.\ Phys.\ }}

\def\PR{{\it Phys.\ Rev.\ }}

\def\np#1#2#3{Nucl. Phys. {\bf #1} (#2) #3}
\def\pl#1#2#3{Phys. Lett. {\bf #1} (#2) #3}

\def\sjnp#1#2#3{Sov. J. Nucl. Phys. {\bf #1} (#2) #3}
\def\blankref#1#2#3{   {\bf #1} (#2) #3}

\def\DATE{February 1992}

\Title{
\vbox{\hbox{SSCL--Preprint--64}
\hbox{BROWN HET--851}}}
{{ Heavy Mesons in Two Dimensions }}

\centerline{Benjam\'\i n Grinstein\authornote{$^\dagger$}%
{On leave of absence from Harvard University. Research supported in part by
the Alfred P. Sloan Foundation and by the
Department of Energy under contract DE--AC35--89ER40486.
\hfill\break
email: grinstein@sscvx1.bitnet,  @sscvx1.ssc.gov}
and Paul F.~Mende\authornote{$^\ddagger$}
{Research supported in part by DOE grant DE-AC02-76-ER03130.
\hfill\break
email: mende@het.brown.edu}}
\smallskip
\centerline{\it $^\dagger$Superconducting Super Collider Laboratory,
Dallas, Texas 75237}
\smallskip
\centerline{\it $^\ddagger$Department of Physics, Brown University,
Providence, Rhode Island 02912}

\ABSTRACT{
The large mass limit of QCD uncovers symmetries that are not present
in the  QCD lagrangian. These symmetries have been applied to physical
(finite mass) systems, such as $B$ and $D$ mesons. We explore the
validity of this approximation in the 't~Hooft model (two dimensional
QCD in the large $N$ approximation). We find that the large mass
approximation is good, even at the charm mass, for form factors, but
it breaks down for the pseudoscalar decay constant.
}


\newsec{Introduction}
Heavy mesons play a prominent role in our understanding of fundamental
processes.
  From their weak decays we may extract fundamental parameters of the
standard model of electroweak interactions.
Rare decays are sensitive to the presence of new fundamental forces.
And they offer the exciting possibility
of observing for the first time violation of CP invariance in a decay process.

Straightforward interpretation of measured lifetimes
and branching fractions is marred by the difficulties that
strong interactions present for practical
calculations. Monte-Carlo  simulations of lattice QCD may eventually
furnish accurate calculations of the matrix elements that
are relevant to these processes.
An alternative approach is furnished by the heavy quark effective
theory (HQET) formalism.
Approximate spin and flavor
symmetries\ref\isgurwise{N. Isgur and M.B. Wise, \pl{B232}{1989}{113};
\pl{B237}{1990}{527}}\ of
the $S$-matrix in the one heavy hadron sector are made explicit.
The symmetries become exact in the limiting case of
infinitely massive  heavy quarks.
  From these symmetries a number of remarkable results follow,
such as the normalization of
form factors for semileptonic $B\to D$ and $D^*$ transitions at maximum
momentum transfer, $q^2_{\rm max}$, and a set of five relations among the six
corresponding form factors which hold at any momentum transfer.

In this regard it is of paramount importance to determine the accuracy of the
large-mass approximation since in reality
the charm and bottom quark masses are
both only  factors of a few larger than, say,  the $\rho$-meson mass.
Unfortunately, this issue involves non-perturbative matrix elements and is
therefore hard to pin down. Moreover, one has to consider each physical
quantity separately, as the approach to the asymptotic regime of
infinite masses may be faster for some than for others.
For example, some form factors in semileptonic decay
remain calculable at $q^2_{\rm max}$
even after  $1/m$ corrections are included
\ref\luke{M.E. Luke, \pl{B252}{1990}{447}};
and Monte-Carlo  simulations of lattice QCD  in the
quenched approximation\ref\chris{
P. Boucaud \etal,
\pl{B220}{1989}{219}\semi
C.R. Allton \etal,
\np{B349}{1991}{598}}\ indicate that~$1/m$ corrections
to the pseudoscalar decay constant of a heavy meson
are large (of the order of 40\% at the charm mass).

In this letter we report on investigations of this issue using two-dimensional
QCD in the $1/N$~expansion as a model of the strong interactions.
It is worth emphasizing from the outset that such model
calculations are uncontrolled
approximations to four-dimensional QCD,
and as such one should refrain from
using the quantitative results as estimates for physical observables.
We believe the same comment applies to {\it any} model calculation,
like the ones
afforded by potential models or QCD sum-rules.
By gathering evidence for
qualitative features that are common to all
physically reasonable models one can
begin to believe that such features are also present in QCD.

An important limitation of this model is the lack of spin,
and we shall have nothing to say about relations that follow from the
spin symmetry of the HQET.
On the other hand, because the model is a relativistic field
theory exhibiting confinement, it is ideally suited to test
the scale of the onset of flavor symmetries.

\OMIT{
The paper is organized as follows.
In section~2 we briefly review some salient features of the model.
In section~3 we compute the pseudoscalar decay constant,
and in section~4 we compute form factors for heavy to heavy transitions.
We close in section~5 with a discussion of these results.
}
In the following sections we review some salient features of the model,
compute the pseudoscalar decay constant,
compute the form factors for heavy to heavy transitions,
and close with a discussion of these results.

\REF\thooft{G.~'t~Hooft,
         \art{A two-dimensional model for mesons,}
        \NP {\bf B75} (1974) 461}

\REF\CCG{C.~G.~Callan, N.~Coote, and D.~J.~Gross,
         \art{Two-dimensional Yang-Mills theory: a model of quark confinement,}
        \PR {\bf D13} (1976) 1649}

\REF\einhorn{M.~B.~Einhorn,
         \art{Confinement, form factors, and deep-inelastic scattering in
        two-dimensional quantum chromodynamics,}
        \PR {\bf D14} (1976) 3451}

\REF\multhopp{
        A.J.~Hanson, R.D.~Peccei, and M.K.~Prasad,
         \art{Two-dimensional SU($N$) gauge theory, strings and wings:
        comparative analysis of meson spectra and covariance,}
        \NP {\bf B121} (1977) 477;
        \nextline
        S.~Huang, J.W.~Negele, and J.~Polonyi,
         \art{Meson structure in QCD$_2$,}
        \NP {\bf B307} (1988) 669 }

\REF\JM{
	R.L.~Jaffe and P.F.~Mende,
	 \art{When is Field Theory Effective?,}
	\NP {\bf B369} (1992) 189 }

\newsec{The 't~Hooft Model}
We calculate the properties of heavy mesons in 1+1 dimensions.
This model of QCD, in the $1/N$ expansion, was solved by 't~Hooft~\thooft .
It shares features of the four-dimensional $1/N$ expansion which,
in turn, has some common ground with meson phenomenology.
The spectrum consists of meson states,
with an approximately linear, Regge-like trajectory.
Since there is no spin in two dimensions, these are obviously radial
excitations --- there is no analogue of the spin-symmetry relations
which appear so fruitful in studies of the real world.
Nor are there baryon or glueball states, these being suppressed by
the $1/N$ expansion and the lack of transverse dimensions.
Yet in spite of the peculiarities of two dimensions, it is nonetheless
a non-trivial strong-coupling solvable theory which is ideally
suited for testing our dynamical ideas.
This model has been extensively studied and we
refer the reader for details of the calcul\-ations to
\refs{\CCG, \einhorn} where the formalism for the matrix elements
was derived, to \multhopp\ where the numerical methods are discussed,
and to \JM, where they were recently applied to the study of low-energy
effective theories.

We take two flavors of quark, one heavy and one light, which we
denote as $Q$ and $q$, respectively, with bare masses $M$ and $m$.
The coupling constant~$g$ has dimensions of mass, and we work in units
where $g^2 N /\pi \equiv 1$.
The lowest heavy-light $Q\bar q$ bound state ---let us call it
the~$B$--- is a pseudoscalar of mass~$\MU$.
We can solve the non-perturbative bound-state
equation numerically for the wave functions and the masses\multhopp.

As our interest is in the approach to asymptotia, we hold $m$ fixed
(taking the value $m^2=0.3$ throughout the paper),
computing the dependence of the universal
form factors and pseudoscalar decay constant, $f_B$, as functions
of~$M$.
\REF\ggfgl{B. Grinstein, \np{B339}{1990}{253}\semi%
H. Georgi, \pl{B240}{1990}{447}\semi%
A. Falk, B. Grinstein and M. Luke, \np{B357}{1991}{185}}
\footnote{*}{It is also possible to take the limit $M\to\infty$ directly
in the field theory before solving for the bound states,
as has been done in 4D to derive the HQET\ggfgl.
We have found the bound state equation appropriate to this
limit, which will be reported elsewhere.}

\newsec {The pseudoscalar decay constant}
The pseudoscalar decay constant $f_B$ for the meson
$B$ is defined by
\eqn\fdefined{
   \vev{0|\bar q \gamma^\mu\gamma^5 Q |B(p)}= f_B p^\mu~.
}
If the states are given the usual relativistic normalization,
\eqn\normalize{
   \vev{B(p')|B(p)}=2E\delta(p-p')~,
}
then the large-mass limit gives the scaling behavior
   $f_B\sqrt{\MU}\sim {\rm constant} .$
This follows from the observation that the static properties
 of a `dressed' heavy source of color are  independent of its mass. The factor
of $\sqrt\MU$ simply reflects the mass-dependent normalization of
states, {\it cf.\/} eq.~\normalize.
In four dimensions there are logarithmic corrections to this
relation\ref\volopol{M.B. Voloshin and M.A. Shifman, \sjnp{45}{1987}{292}\semi
H.D. Politzer and M.B. Wise, \pl{B206}{1988}{681}; \blankref{B208}{1988}{504}}.
We have omitted these foreseeing that they are absent
in a super-renormalizable theory.

In two dimensions both vector and axial currents are good interpolating
fields for the pseudoscalar meson. Since $\gamma^5\gamma^\mu=\epsilon^{\mu\nu}
\gamma_\nu$, they  are both characterized by the same decay constant.
\OMIT{
One has
\eqn\ftwodefined{
   \vev{0|\bar q \gamma_\mu\gamma^5 Q |B(p)}= \epsilon_{\mu\nu}
   \vev{0|\bar q \gamma^\nu Q |B(p)}= f_B p_\mu~.
}
}
In the 't~Hooft model $f_B$ is easily computed\refs{\CCG, \einhorn}. It is
given by\footnote{$^\star$}{We drop
an $N$-dependent prefactor which cancels out
of observables.}
 $f_B = \int_0^1 \dt\,\phi_B(t)$,
where $\phi_B(t)$ is the momentum-space wave function
for the $B$-meson and~$t$ is the fraction of light-cone momentum
carried by the heavy quark~\thooft.
In  \fig\decayc{Plot of the pseudoscalar decay constant $f_B\sqrt{\MU}$
as a function of the mass of the heavy quark.
The mass of the  light quark, $m$, is fixed throughout: $m^2=0.3$.}\ we
show how the limiting behavior $f_B\sqrt{\MU}\sim {\rm constant}$ is attained.

Fitting the high-mass portion of the curve ($M\ge 5$) to a
quadratic polynomial in~$1/M$ gives a description of the $1/M$ corrections:
\eqn\fMfit{
f_B\sqrt{\MU}= 2.0\left[1-{1.4\over M} + \left({1.4\over
M}\right)^2\right]~. }
There are two sources of uncertainty in this calculation:
the endpoint fit of the wavefunctions $\phi_B$
and the few values of the mass we take for our mass fit.
By varying the number of sampling points and successively
improving the quality of
our wavefunctions we estimate the error to
be less than~2\% for the coefficient
of~$1/M$ and less than~20\% for that of~$1/M^2$.
\INSERTFIG{
\includegraphics{HM1.ps}
\vglue2.5truein
{\ninerm
Figure 1.
{\ninesl Plot of the pseudoscalar decay constant $f_B\sqrt{\MU}$
as a function of the mass of the heavy quark.
The mass of the  light quark, $m$, is fixed throughout: $m^2=0.3$.}
\medskip}
}

\newsec{Form factors}
Of greatest interest in probing the structure of the theory in the
heavy-quark limit is the
form factor for a heavy-quark current between heavy mesons:
\eqn\ffsdefined{
   \vev{B'(p')|\bar Q' \gamma_\mu Q |B(p)}=
   f_+(q^2)(p+p')_\mu + f_-(q^2)(p-p')_\mu~,
}
where the momentum transfer is given by $q=p-p'$.
$B$ and $B'$ have the same light quark content but different
heavy quarks.
When \OMIT{the heavy quarks are identical,} $Q'=Q$,
\OMIT{and the current does not
change flavors, }
conservation of current gives $f_-=0$.
The remaining form factor, \OMIT{in that case,
denoted by} $f(q^2)$, is normalized,  $f(0)$=1.

In the limit of infinite masses,
the mesons are more appropriately labelled by
their velocities $v=p/\MU$ and $v'=p'/{\MU'}$,
and it is natural to describe the dependence on~$q^2$ through
the function
$w \equiv \vv = (\MU^2+\MU^{\prime2}-q^2) / 2\MU\MU'$.
The transition amplitude for a `dressed'
heavy source of color with velocity $v$ to a second one with velocity
$v'$ is then independent of their masses.
Thus all three form factors in this limit
are given in terms of a single  universal `Isgur-Wise'
function\isgurwise, $\xi(\vv)$: \OMIT{, known as the Isgur-Wise function.
One obtains}
\eqn\iwffs{
f_\pm(q^2) =\xi(\vv)\left({\MU'\pm \MU\over2\sqrt{\MU'\MU}}\right)~,
\quad
f(q^2) = \xi(\vv)~.
}
As before, we ignore
logarithmic corrections\ref\fggw{A. Falk \etal,
\np{B343}{1990}{1}}\ which are absent in the super-renormaliz\-able
model.

The Isgur-Wise function is normalized ---$\xi(1)=1$--- by
evaluating it for identical heavy mesons so that
$\vv=1$ corresponds to~$q^2=0$.
This in turn gives a prediction in the case of different heavy mesons
\OMIT{for then $\vv=1$ corresponds to} at $q^2=q^2_{\rm max}=(\MU-\MU')^2$:
\OMIT{the maximum physical momentum transfer:}
\eqn\iwffsnormalized{
f_\pm(q^2_{\rm max}) =\left({{\MU'}\pm \MU\over2\sqrt{\MU'\MU}}\right)~.
}

Let us now examine the behavior of the form factors for {\it finite mass\/}
heavy quarks in two dimensions, where they may be computed exactly.
In two and four dimensions the definitions and infinite mass relations
are identical.
\OMIT{
We would like to see first, how the Isgur-Wise function behaves
and what heavy-meson relations may be derived in this case; and second,
how the limit of infinite mass is approached.
}
What are the dominant~$1/M$ corrections and where do they set in?
How well can the form factors be approximated from the quark model
contributions alone?

It is straightforward to evaluate the current matrix elements of interest.
For spacelike momentum transfer the minus component of the current
gives the following expression\einhorn \ in which the first term dominates
our calculation:
\eqnn\einhornff
$$
   \eqalign{
&   \vev{B'|V_-|B}/2q_- =
   \int_1^\omega \dt\, \phi_B\left({t\over\omega}\right)
   \phi_{B'}\left({1-t\over1-\omega}\right)
   \cr & \qquad
    +{1\over 1 - \omega }
   \int_0^1 \dt\, \phi_B\left({t\over\omega}\right)
   \Phi_{B'}\left({1-t\over1-\omega}\right) G(t;q^2)
   \cr & \qquad\qquad
   +\int_1^\omega \dt\, \phi_B\left({t\over\omega}\right)
   \phi_{B'}\left({1-t\over1-\omega}\right) {\tilde G}(t;q^2)\cr
}
\eqno\einhornff
$$
where $\omega \equiv p_-/q_-$,
the full vertex $\Phi(t)\equiv\pvint_0^1\dt'\,\phi(t')/(t'-t)^2$,
the Green function
$G(t;q^2)\equiv \sum_n f_n\phi^{QQ'}_n(t)/(q^2-\mu_n^2)$,
and
${\tilde G}(t;q^2)\equiv\pvint\dt'\,G(t';q^2)/(t-t')^2$.

The first line represents the `quark-model' contribution to the form factor,
where the current couples directly to the valence heavy quark.
The other terms represent the full set of
remaining graphs which arise from
gluon exchange in the current channel.
These can be resummed explicitly and computed; we have done
this and checked that they give but small contributions for the
cases we studied.

First we consider the form factor $f(q^2)$ for identical heavy mesons,
$B'=B$.
In \fig\formfs{Heavy-quark current form factors $f(\vv)$
functions of $\vv$ for different masses, $M$, of the heavy quark.
 The solid, dashed and dashed-dotted
lines correspond to $M=450$ (effectively infinite), 14 and 5,
respectively.} we plot the heavy-quark current form factors
for $M=5$, 14 and 450. \OMIT{different heavy quark masses. }
\OMIT{The solid, dashed and dashed-dotted
lines correspond to $M= 450$ (effectively infinite),
14 and 5, respectively.}
\INSERTFIG{
\includegraphics{HM2.ps}
\vglue2.5truein
{\ninerm
Figure 2.
{\ninesl
Heavy-quark current form factors $f(\vv)$
functions of $\vv$ for different masses, $M$, of the heavy quark.
 The solid, dashed and dashed-dotted
lines correspond to $M=450$ (effectively infinite), 14 and 5,
respectively.}
\medskip}
}
To better characterize the large-mass behavior we can examine,
at fixed~$w=\vv$, the approach to the asymptotic function
as the heavy quark mass is varied away from $M=\infty$.
We calculated the form factors for additional masses,
$M=45$ and~140. A fit to the quadratic polynomial in~$1/M'$,
\eqn\quadfitffs{
   f(\vv) = \xi(\vv) \left[1 - {\kappa_1(\vv)\over M'} -
   {\kappa_2(\vv)\over {M'}^2} \right] ~~,
}
gives $\kappa_1=0.10$, 0.26 and 0.32, and $\kappa_2=0.6$, 1.6 and 3.0, for
$\vv=1.25$, 2.0 and 3.5, respectively. As was the case for the pseudoscalar
decay constant, these quantities are good to about 10\% accuracy.

Next we consider transitions between  different ground state mesons.
At $q^2_{\rm max}$ the left hand side of \einhornff\ is predicted
 in the HQET, {\it cf.\/} eqs.~\ffsdefined\ and \iwffsnormalized:
\eqn\difflavors{
   \left.{\vev{B'(p')|\bar Q' \gamma_- Q |B(p)}}
   \right|_{q^2_{\rm max}} =
   {2\sqrt{\MU\MU'}}
}

In \fig\ffsdifmasses{Difference $d(M')$ from unity of the matrix element
of the flavor changing current
between ground state mesons, at $q^2_{\rm max}$, normalized to
$2\sqrt{\MU\MU'}$,
as a function of the heavy quark mass
$M'$ in the lighter meson; see eq.~\noname.
The heavier meson has $M=450$.}\ we show
\eqn\noname{
   d(M') \equiv
   {\left.{\vev{B'(p')|\bar Q' \gamma_- Q |B(p)}  }
   \right|_{q^2_{\rm max}} \over
   2\sqrt{\MU\MU'}      
   }-1
}
as a function of the heavy quark mass~$M'$ in the lighter $B'$-meson.
\INSERTFIG{
\includegraphics{HM3.ps}
\vglue2.5truein
{\ninerm
Figure 3.
{\ninesl
Difference $d(M')$ from unity of the matrix element
of the flavor changing current
between ground state mesons, at $q^2_{\rm max}$, normalized to
$2\sqrt{\MU\MU'}$,
as a function of the heavy quark mass
$M'$ in the lighter meson; see eq.~\noname.
}\medskip}
}
The heavier $B$~meson has heavy quark mass
held fixed at~$M=450$.
This plot reveals that the finite-mass corrections are {\it quadratic\/}
in~$1/M'$.
This result is expected as a consequence of Luke's
theorem\luke, which states that there are no corrections of order $1/M'$ to
the predicted normalization of form factors at maximum momentum
transfer; corrections begin at order~$1/{M'}^2$
\footnote{*}
{\OMIT{Actually, a stronger version of Luke's theorem is given by
Cho and Grinstein (to be published).} Luke's theorem, in its original version,
ignores corrections of order $\bar\alpha_s(M')/M'$. A stronger version
(Cho and Grinstein, to be published) gives
the absence of corrections of order $1/M'$ to
all orders in the strong coupling.
}.
Again the result can be fit to a quadratic polynomial in $1/M'$:
\eqn\lastfit{
   d(M') =
   - {0.002\pm0.002\over M'} - {0.14\pm0.01\over M^{\prime2}~.}
}
Here we have written the uncertainty as a range
since the first term is consistent with zero.
Alternatively, one can fit to a power $A/M^{\prime n}$, for
which we find $A=-0.17$ and $n=2.1$,
which clearly indicates the absence of
$1/M'$ corrections.

\newsec{Discussion and Outlook}
\OMIT{
We see there can be little doubt of the validity of the large-mass expansion.
The important question is to determine the mass
at which the approximation is already reasonably good. There are
indications in Monte-Carlo simulations of lattice QCD in the
quenched approximation that the HQET is a poor approximation
at the charm mass\chris. Is this generally true?
}
\OMIT{
A glance at the results of eqs.~\fMfit, \quadfitffs\ and \lastfit\
demonstrates that this is not the case.
}
The size of ~$1/M$ corrections is not uniform and cannot be characterized
by $f_B$ alone:
they are far larger for the pseudoscalar
decay constant than they are for the form factors,
for which the asymptotic limit is approached rapidly.
This is true both for the functional form of the
form factors out to moderate $\vv$ (eq.~\quadfitffs) as well as for the
normalization of the form factors of flavor changing-currents (eq.~\lastfit).
A similar pattern of corrections is observed in calculations that use
the non-relativistic potential
model of quarks\ref\private{N. Isgur, private communication,
unpublished.}.

As we explained earlier, we do not view the model as a means of extracting
quantitative information for four dimensions.
Nevertheless the question of whether the
large mass limit applies for the charm quark can be addressed.
\OMIT{
How do we identify the relevant scale,
the ``charm quark mass'' in this model?
A heavy quark~$Q$ is one for which $Q\bar Q$ bound states are
Coulombic, or ``onium''-like; and
the charm mass is, in a  sense, the lightest mass a heavy quark can have.
}
The transition from light to heavy quark dynamics was explored
in detail in ref.~\JM.
Inspection of the spectrum of $Q\bar Q$
states and the strength of the singularities in the form
factors reveals this mass to be
between~1 and~2 (again in units of $g^2N/\pi=1$)~\JM.
For this \OMIT{lowest
possible heavy quark} mass the 't~Hooft model has corrections of order
100\% to the HQET prediction of the pseudoscalar decay constant, while
it gives small corrections, 4--14\%, to the normalization of form factors of
flavor changing currents at $\vv=1$.

It is apparent that the 't~Hooft model is a valuable testing ground for
our ideas of the large mass limit of QCD.
\OMIT{We hope to apply it in the future to
other interesting problems. For example,}
One may attempt to investigate the
validity of Bjorken's sum
rule\ref\bjorken{J.D. Bjorken, \OMIT{
talk presented at Les Recontres de Physique de
la Vallee d'Acoste, La Thuile, Italy, March 18-24, 1990,} SLAC-PUB-5278
(1990) }, since one can compute the
wavefunctions of rather high excited states of $Q\bar q$ mesons.
More interestingly, one can address the
question of whether the form factor for the
semileptonic decay of a heavy meson
into a light one is wave-function or pole dominated.
The answer remains elusive due to the intrinsically
non-perturbative nature of dynamics involved. \OMIT{The
't~Hooft model seems well-suited to address this issue.}

\medskip\noindent
{\it Note added --- }
After completing this work we learned about similar calculations
in progress by   M. Burkardt and E. Swanson.


\smallskip
\hbox to \hsize{\hss\vrule width2.5cm height1pt \hss}
\smallskip

{\ninerm
$^\dagger$On leave of absence from Harvard University.
Research supported in part by  the Alfred P. Sloan  Foundation and by the
 Department of Energy under contract DE--AC35--89ER40486.
email: grinstein@sscvx1.bitnet,  @sscvx1.ssc.gov.
\hfill\break
$^\ddagger$Research supported in part by DOE grant DE-AC02-76-ER03130.
email: mende@het.brown.edu
\hfill\break
 \immediate\closeout\rfile \input refs.tmp 
}
\enddoublecolumns
\output{\onepageout{\unvbox255}}
\vfill\eject\end